\documentclass[aps,prl,twocolumn,showpacs,amsmath,amssymb,superscriptaddress,floatfix]{revtex4-1}

\usepackage[dvipsnames]{xcolor}
\usepackage{graphics}      
\usepackage[pdftex]{graphicx}      
\usepackage{bm}
\usepackage{natbib}            
\usepackage[colorlinks,plainpages=false,linkcolor=blue,urlcolor=blue,citecolor=blue,pdfpagemode=UseNone,pdfstartview=FitBH]{hyperref}
\usepackage{comment}

\usepackage{ifpdf}
\ifpdf

\begin{document}

\title{Increasing the number of topological nodal lines in semimetals via uniaxial pressure}

\author{Adolfo O. Fumega}
  \email{adolfo.otero.fumega@usc.es}
 \affiliation{Departamento de F\'{i}sica Aplicada,
  Universidade de Santiago de Compostela, E-15782 Campus Sur s/n,
  Santiago de Compostela, Spain}
\affiliation{Instituto de Investigaci\'{o}ns Tecnol\'{o}xicas,
  Universidade de Santiago de Compostela, E-15782 Campus Sur s/n,
  Santiago de Compostela, Spain} 
\author{Victor Pardo}
\affiliation{Departamento de F\'{i}sica Aplicada,
  Universidade de Santiago de Compostela, E-15782 Campus Sur s/n,
  Santiago de Compostela, Spain}
\affiliation{Instituto de Investigaci\'{o}ns Tecnol\'{o}xicas,
  Universidade de Santiago de Compostela, E-15782 Campus Sur s/n,
  Santiago de Compostela, Spain} 
  \author{A. Cortijo}
    \email{alberto.cortijo@uam.es}
  \affiliation{Departamento de F\'isica de la Materia Condensada, Universidad Aut\'onoma de Madrid, Madrid E-28049, Spain} 
  
\begin{abstract} 
The application of pressure has been demonstrated to induce intriguing phase transitions in topological nodal-line semimetals. In this work we analyze how uniaxial pressure affects the topological character of BaSn$_2$, a Dirac nodal-line semimetal in the absence of spin-orbit coupling. Using calculations based on the density functional theory and a model tight-binding Hamiltonian, we find the emergence of a second nodal line for pressures higher than 4 GPa. We examine the topological features of both phases demonstrating that a nontrivial character is present in both of them. Thus, providing evidence of a topological-to-topological phase transition in which the number of topological nodal lines increases. The orbital overlap increase between Ba $d_{xz}$ and $d_{yz}$ orbitals and Sn $p_z$ orbitals and the preservation of crystal symmetries are found to be responsible for the advent of this transition. Furthermore, we pave the way to experimentally test this kind of transition by obtaining a topological relation between the zero-energy modes that arise in each phase when a magnetic field is applied.

\end{abstract}

\maketitle

\section{Introduction}

In the last decades the concept of topology has provided a different perspective to look at materials. 
This field was initiated with the discovery of topological insulators \cite{review_TI_kane} followed by topological semimetals \cite{review_TSM_Mele}. 
Inside the latter category, topological nodal-line semimetals (NLSM) are three dimensional systems in which the valence and conduction bands near the Fermi level cross each other forming a one-dimensional loop in momentum space \cite{svo_sto_pickett,PhysRevB.84.235126}. These crossings produce drumheadlike surface states, which prompt the possibility to study strongly correlated surface phases such as superconductivity or ferromagnetism \cite{PhysRevB.83.220503, PhysRevLett.122.016803, Shao2020}. Moreover, topological NLSM have been recently stated to occur also in two-dimensional nitrogen-based systems  \cite{PhysRevMaterials.3.084201,PhysRevB.102.075133,PhysRevB.102.125118}.


Three dimensional topological nodal lines have been reported in the family of ZrSiX materials (where X corresponds to a chalcogen atom) \cite{Schoop2016, PhysRevLett.117.016602, PhysRevB.93.201104}.
In recent studies, it has been shown that pressure can induce topological phase transitions in these nodal-line semimetals. 
On the one hand, in the case of ZrSiS pressure induces a topological nontrivial-to-trivial transition \cite{PhysRevB.99.085204}. It has been attributed to a weak lattice distortion by nonhydrostatic compression in which crystal symmetries are not preserved \cite{PhysRevB.100.205124}.
On the other hand, infrared spectroscopy experiments have suggested that ZrSiTe undergoes two phase transitions at 4.1 and 6.5 GPa \cite{PhysRevB.99.245133}. Raman spectroscopy and X-ray diffraction measurements as well as calculations have found that they correspond to Lifshitz transitions, in which the topology of the Fermi surface changes without breaking any lattice symmetry. Interlayer interactions caused by pressure are found to be responsible of these transitions \cite{PhysRevB.101.081108}. However, the topological character of these transitions cannot be elucidated from these experiments. Another question is whether similar transitions appear in other materials hosting NLSM, or instead pressure-induced topological transitions always go in the same direction, from topologically non-trivial to trivial as pressure increases.

\begin{figure}[!h]
  \centering
  \includegraphics[width=\columnwidth]{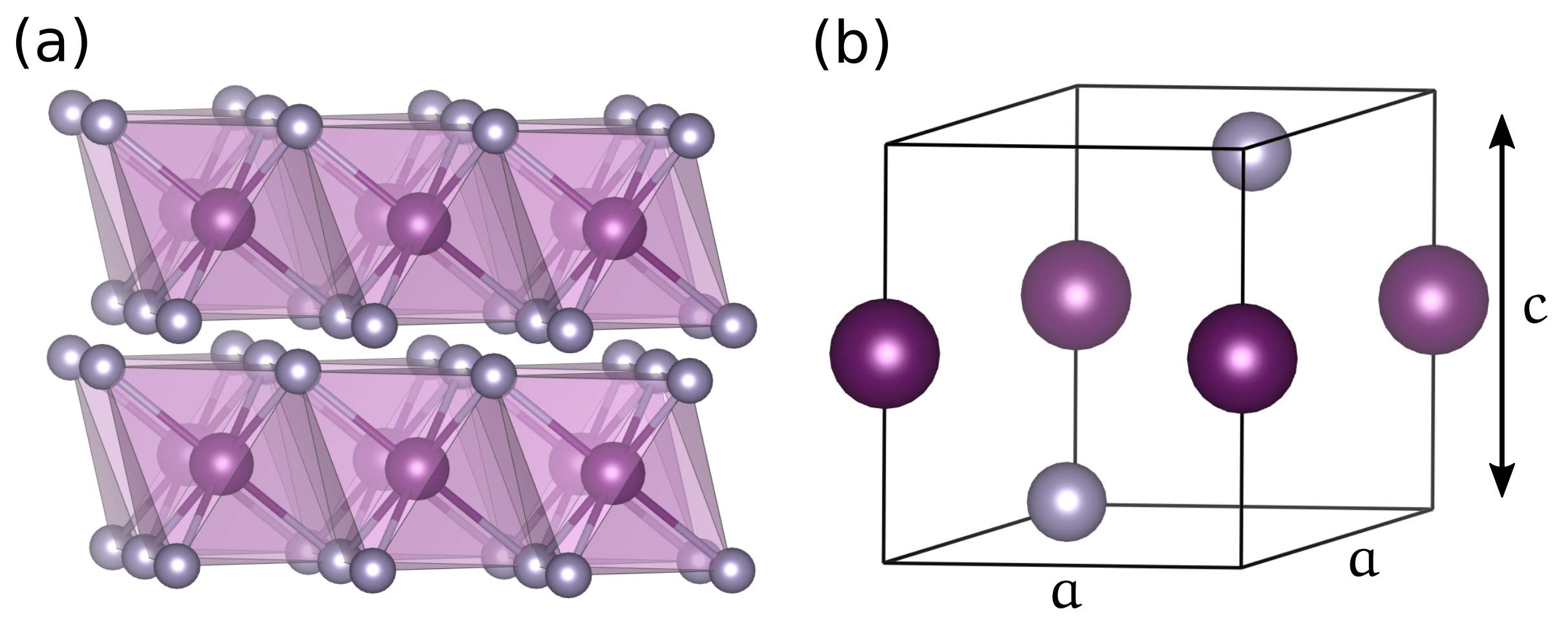}
     \caption{Crystal structure of BaSn$_2$, Ba (Sn) atoms in purple (grey). (a) Side view of the layered structure of BaSn$_2$, where Sn atoms octahedrally coordinate the Ba cations. (b) Unit cell of BaSn$_2$.  Uniaxial pressure is applied along the depicted off-plane $c$ direction. }\label{struct}
\end{figure}

BaSn$_2$ has been reported to be a strong topological insulator with a bulk band gap of about 200 meV \cite{PhysRevB.95.085116}. However, in the absence of spin-orbit coupling (SOC) the system becomes a Dirac nodal-line semimetal \cite{PhysRevB.93.201114}. It is a layered material that crystallizes in the space group no. 164 (Fig. \ref{struct}a). Due to this layered structure, it is expected that BaSn$_2$ is a representative material different from the ZrSiX family to analyze possible pressure-induced topological transitions.

In this work, we study the effect of applying uniaxial pressure to BaSn$_2$ in the absence of SOC. We have found that at around 4 GPa a Lifshitz transition occurs and a second topological nodal line emerges. We provide a microscopic insight to explain this transition and analyze how it affects the topological properties of BaSn$_2$. We also examine a route to experimentally test this kind of novel topological-to-topological transition. Although BaSn$_2$ displays a non negligible SOC coupling in reality, BaSn$_{2}$ in absence of SOC constitutes an ideal system to analyze the possibility of pressure-induced topological phase transitions from a theoretical point of view, as it offer some advantages compared to more realistic NLSMs: is a layered nodal line semimetal at ambient pressure, and the effects of pressure can be easily analyzed. Also, it is a well known system in basis of \emph{ab initio} calculations\cite{PhysRevB.93.201114}. For these reasons, we will use BaSn$_{2}$ as a model system to discuss possible scenarios of topological phase transitions driven by pressure.

\section{Results and discussion} 

We start our analysis by performing first principles electronic structure Density Functional Theory (DFT) calculations\cite{HK,KS} on BaSn$_2$ using an all-electron full potential code ({\sc wien2k}) \cite{WIEN2k}.
The exchange-correlation term was the generalized gradient approximation (GGA) in the Perdew-Burke-Ernzerhof \cite{PBE} scheme, which properly describes the electronic structure of this system\cite{PhysRevB.95.085116,PhysRevB.93.201114}.
The calculations were carried out with a converged \textit{k}-mesh and a value of \textit{R}$_{mt}$\textit{K}$_{max}$= 7.0 and a \textit{R}$_{mt}$ value of 2.5 a. u. for both Ba and Sn. 
Uniaxial pressure was imposed by fixing the experimental \cite{CrystalstructureofbariumdistannideBaSn2} lattice parameter $a=4.652$ \AA$ $ constant and compressing the $c$ lattice parameter (Fig. \ref{struct}b). Energy calculations allow us to obtain $c$ as a function of pressure $P$ given in GPa, $c=5.717-0.136P$ \AA$ $, see Supplementary material. The crystal symmetry is fixed for all the calculations.

\begin{figure}[!h]
  \centering
  \includegraphics[width=\columnwidth]%
    {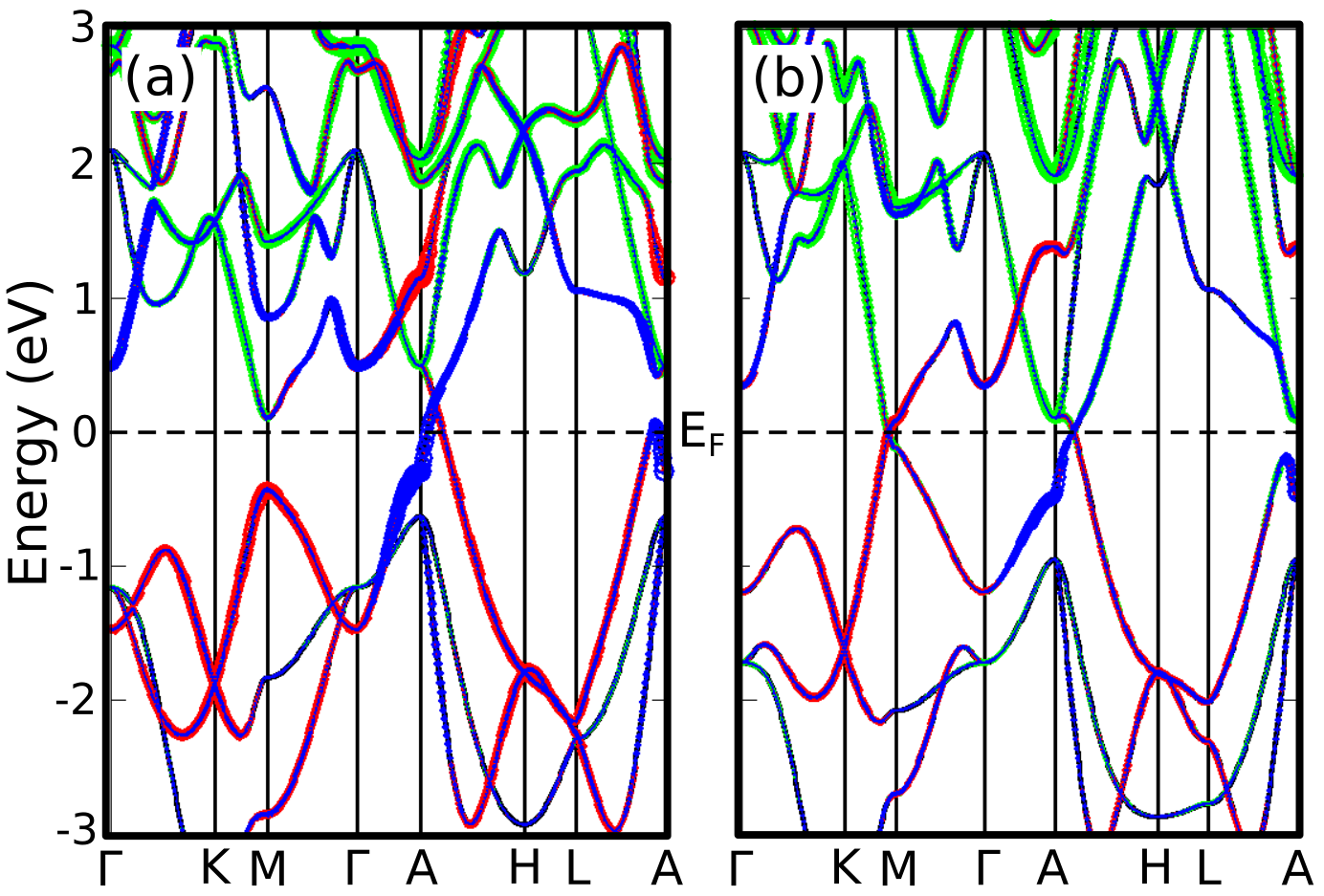}
     \caption{(Color online) DFT band structures for BaSn$_2$. The orbital character is highlighted in colors: Sn $s$ in blue, Sn $p_z$ in red and Ba $d$ in green. The Fermi level is set at zero. (a) Band structure for $P=0$ GPa (T1), there is one band crossing point in the A-H line. (b) Band structure for $P=6$ GPa (T2), a second band crossing appears near the M point.}\label{bandsdft}
\end{figure}

\begin{figure}[!h]
  \centering
  \includegraphics[width=\columnwidth]%
    {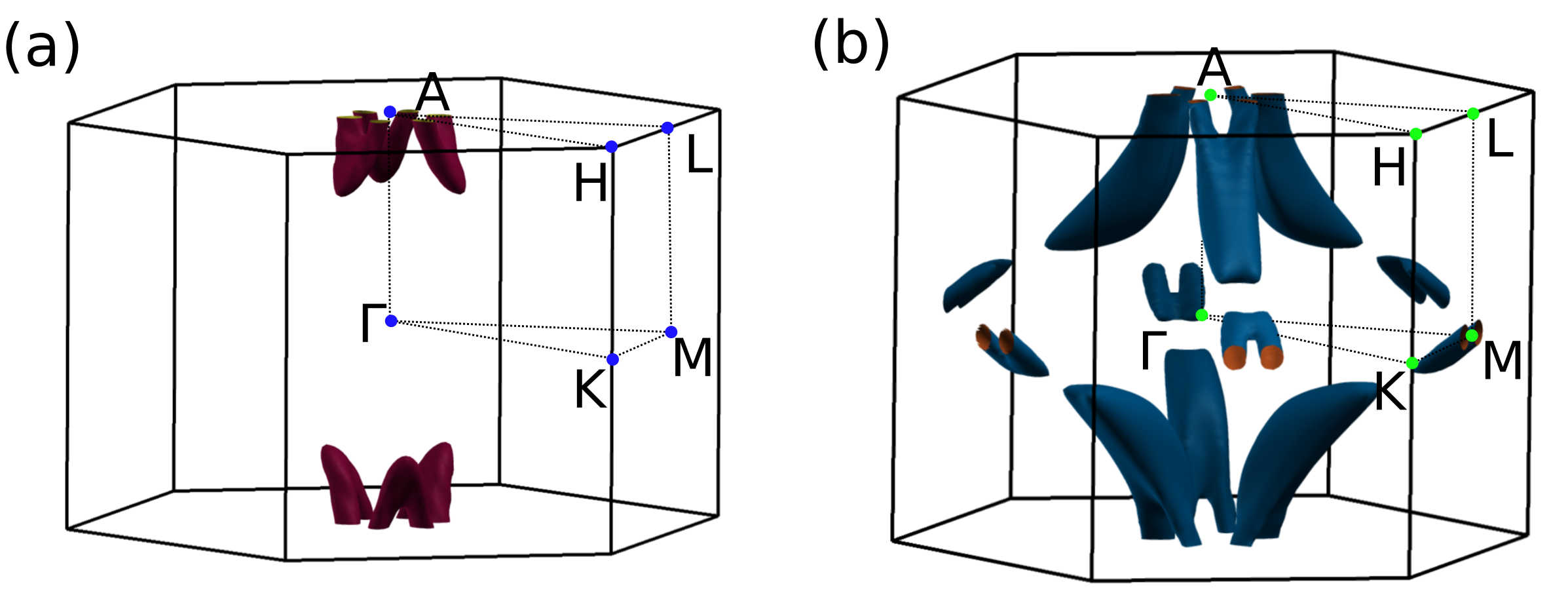}
     \caption{DFT Fermi surfaces for BaSn$_2$: (a) for $P=0$ GPa (T1), it presents a snake-like nodal line around the A point, (b) for $P=6$ GPa (T2), new nodal lines emerge around the M points.}\label{fsdft}
\end{figure}

Figure \ref{bandsdft} shows the DFT band structures of BaSn$_2$ for two different pressure regimes (SOC is not included). The orbital character of the bands is shown as different colors. At low pressure (Fig. \ref{bandsdft}a) one band crossing appears in the A-H line. We can see that a band inversion occurs between $p_z$ (in red) and s (in blue) orbitals at that crossing point. The Fermi surface (Fig. \ref{fsdft}a) shows that this crossing point corresponds to a topological snake-like nodal line around the A point, with an axis parallel to the z-direction. These main features of the low-pressure phase, that we will call T1, have already been studied in detail in Ref. \cite{PhysRevB.93.201114}. The Ba $d$ orbitals (in green in the band structure) are fully unoccupied.

We now turn uniaxial pressure on by reducing the $c$ lattice parameter. The effect in the band structure is that a second crossing point appears along the K-M symmetry line (Fig. \ref{bandsdft}b). The orbital character of the bands involved in this second crossing is Sn $p_z$ and Ba $d$, which suggests that a transition to a different phase (T2) might be taking place. The calculation of the Fermi surface (Fig. \ref{fsdft}b) shows that this new crossing corresponds to a nodal line around the M point. The plane where this new nodal line lives is leaned 45$^{\circ}$ with respect to the z-direction. Since the M point has three arms, three new nodal lines emerge in this new T2 phase.

Note that when SOC is included in these calculations, band crossings are opened and lead to topological insulating states in both phases (see Supplementary material). 

\begin{figure}[!h]
  \centering
  \includegraphics[width=\columnwidth]%
    {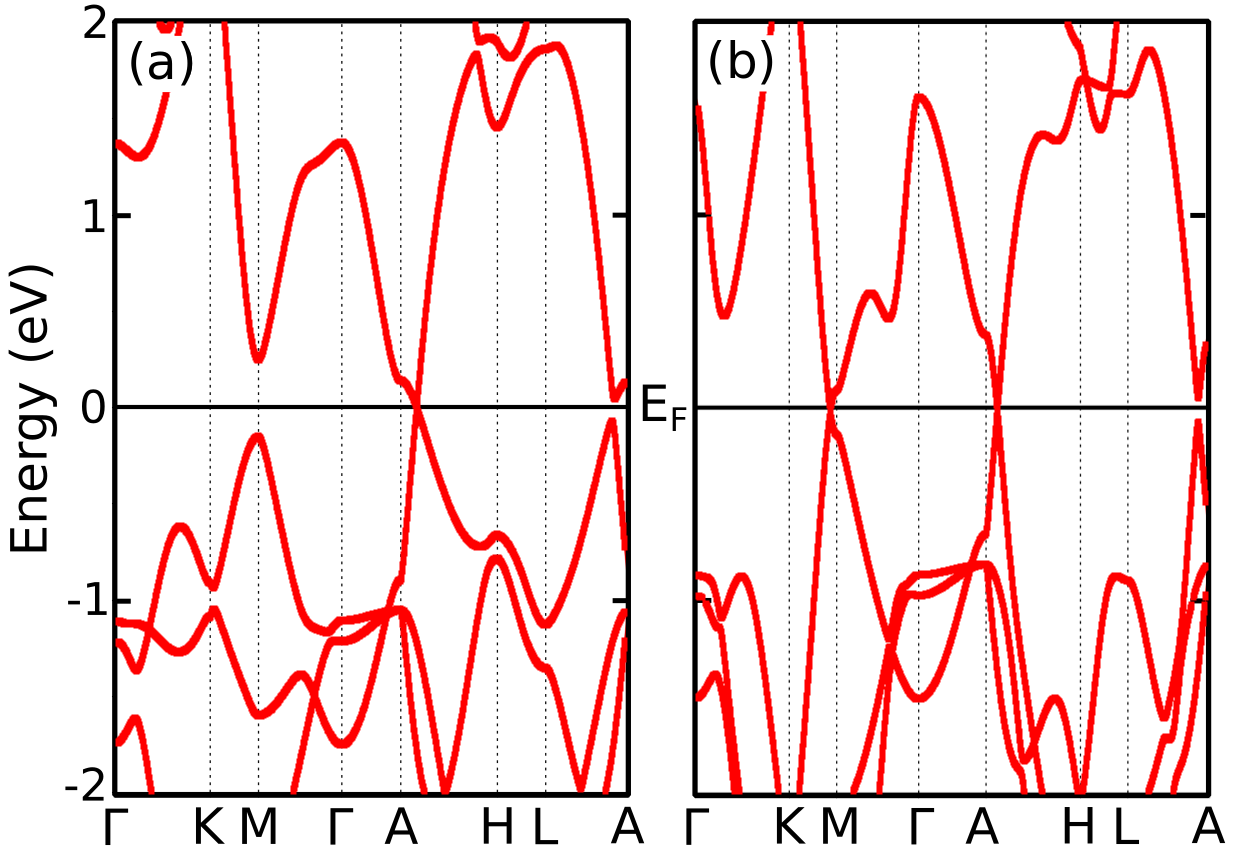}
     \caption{Tight-binding model band structures for BaSn$_2$ (a) for P=0 GPa (T1), the first band crossing point at the A-H line is well reproduced by the model, (b)  for P=6 GPa (T2), the second band crossing near the M point is also reproduced.}\label{bandstb}
\end{figure}

In order to study the topological features and the microscopic origin of the transition, we have constructed a tight-binding (TB) Hamiltonian as a function of uniaxial pressure. It comprises the $s$ and $p$ orbitals of the two Sn atoms in the unit cell and the $d$ orbitals of the Ba atom (thirteen orbitals). Details about the hopping parameters and on-site energies can be seen in the Supplementary material. 
We have used the \texttt{PythTB} \cite{vanderbilt_2018} and \texttt{Pybinding} \cite{moldovan_dean_2017_826942} packages to perform the calculations with our TB Hamiltonian.
In this TB model the only parameters that depend on pressure are the hopping energies between the Ba $d_{xz}$ and  $d_{yz}$ with the Sn $p_z$ orbitals as $t_{d_{xz}p_z}=t_{d_{yz}p_z}=0.7+0.2P$, $P$ being the pressure expressed in GPa and the hopping parameters in $eV$. Therefore, the phase transition is thus governed by this particular hopping.
It has been previously reported that an increase of interlayer interactions without breaking any crystal symmetry might be responsible of the appearance of Lifshitz transitions in ZiSiTe \cite{PhysRevB.101.081108}. Here we find that pressure changes hopping processes within the same Sn-Ba-Sn layer leading to the appearance of a new topological nodal line, without breaking the PT symmetry responsible of the stability of the nodal lines. This is in contrast to the case of ZrSiS, where it has been shown that a topological nontrivial-to-trivial phase transition takes place due to a crystal symmetry breaking\cite{PhysRevB.100.205124}.

\begin{figure}[!h]
  \centering
  \includegraphics[width=\columnwidth]%
    {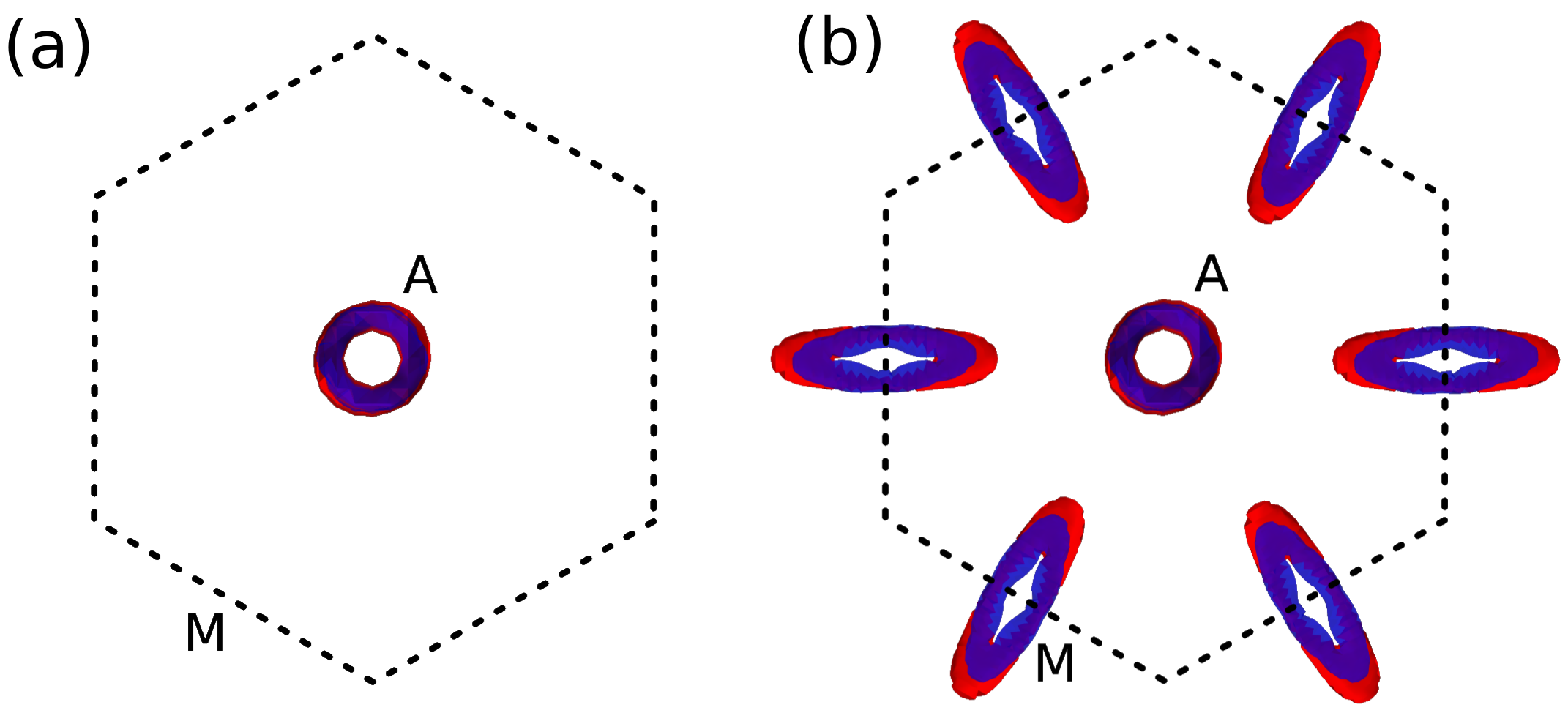}
     \caption{Top view of the TB band touchings within the first Brillouin zone at the Fermi level. The valence (conduction) band is shown in red (blue): (a) for P=0 GPa (T1), the snake-like nodal line is formed around the A point, (b) for P=6 GPa (T2), new nodal lines appear around the M points.}\label{bts_tb}
\end{figure}

Figure \ref{bandstb} shows the computed TB band structures for two given pressures below ($P=0$ GPa) and above ($P=6$ GPa) the transition. The constructed TB model reproduces fairly well the obtained DFT band structures close to the Fermi level in both phases. The analysis of the TB band crossings is shown in Fig. \ref{bts_tb}. It can be seen that, in agreement with the DFT calculations, the TB Hamiltonian correctly predicts a snake-like nodal line around the A point for low pressure. For pressures greater than 4 GPa new nodal lines arise around the M points. Again, these are leaned  45$^{\circ}$ with respect to the z-direction.

\begin{figure}[!h]
  \centering
  \includegraphics[width=\columnwidth]%
    {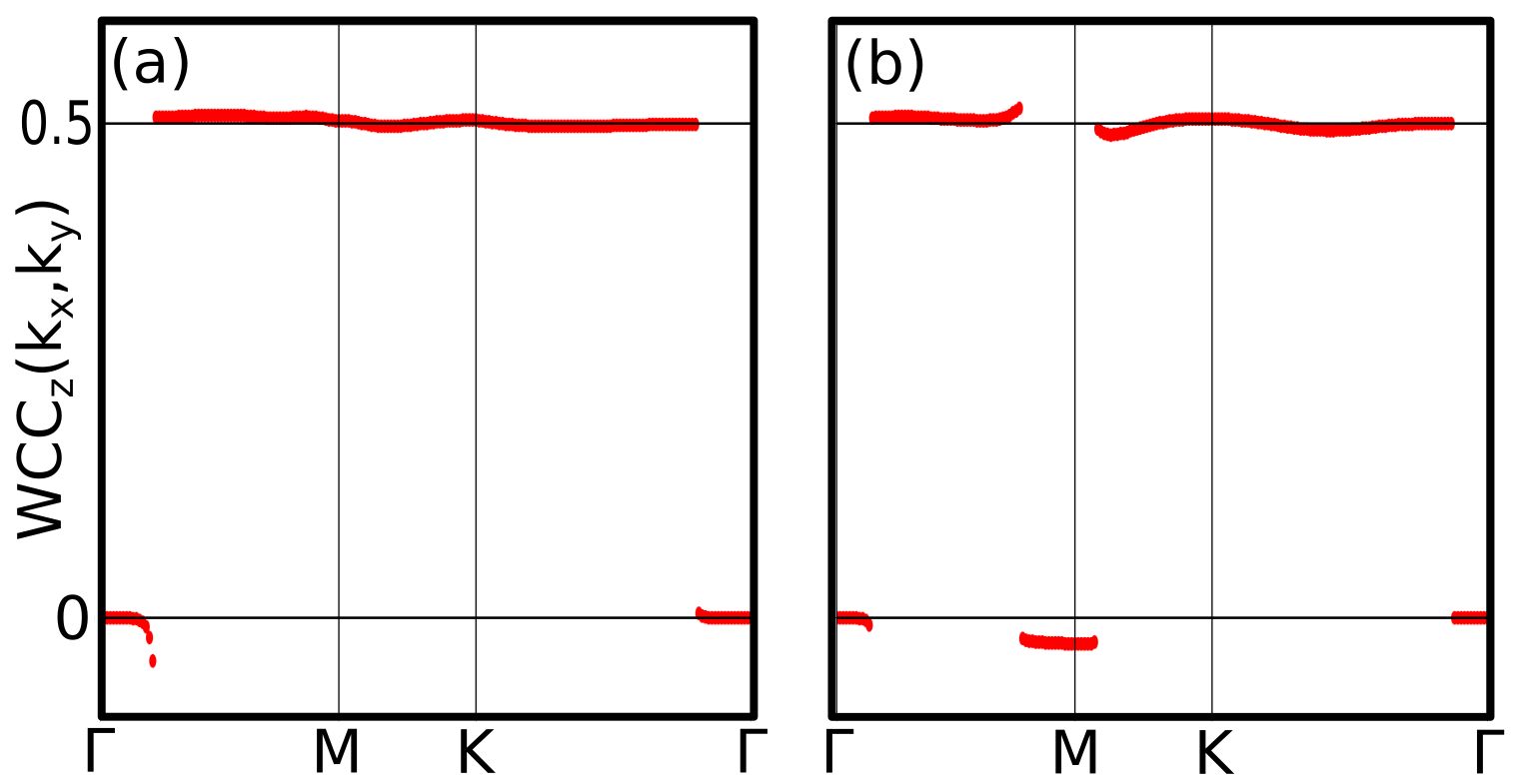}
     \caption{Wannier charge centers (a) For P=0 GPa (T1), around the $\Gamma$ point and due to the snake-like nodal line a jump of 0.5 happens. (b) For P=6 GPa (T2), due to the new topological nodal line a second jump of -0.5 appears around the $M$ point.}\label{wcc}
\end{figure}

\begin{figure}[!h]
  \centering
  \includegraphics[width=\columnwidth]%
    {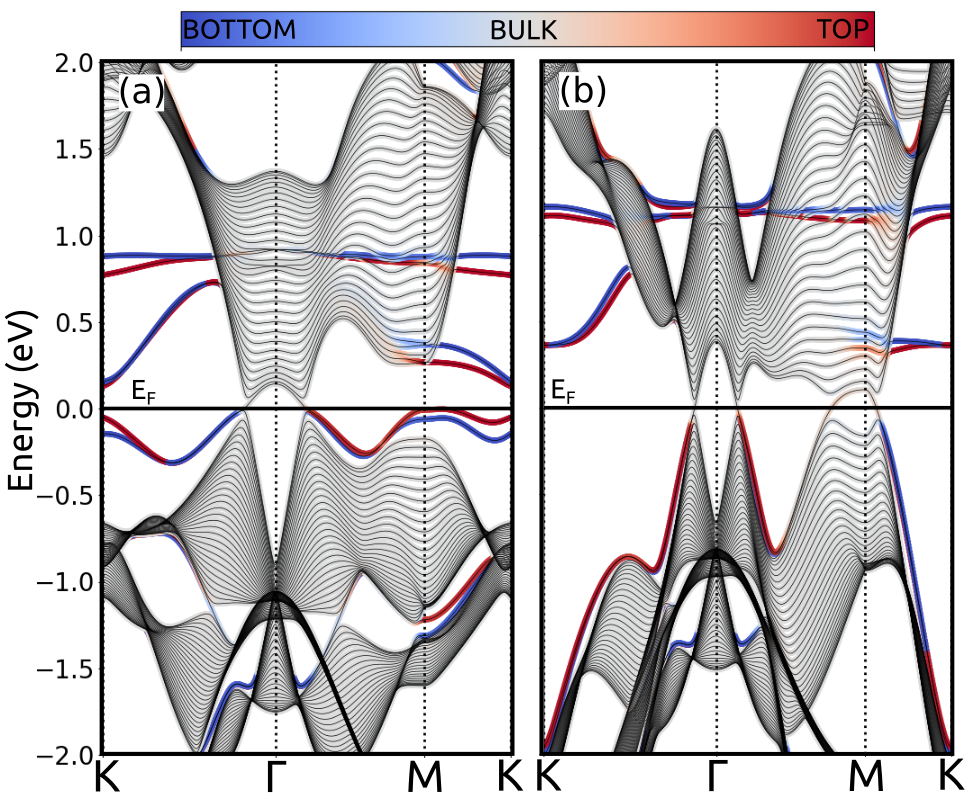}
     \caption{Energy states in a 50-layer structure stacked along the z-direction (Sn-dangling bond terminated). The colormap represents the expectation value of the position operator along the z-direction, in blue (red) the bottom (top) layer and in grey the bulk states. (a) For P=0 (T1), the surface states fill the region outside the projected nodal line around the $\Gamma$ point (b) For P=6 GPa (T2), the surface states get opened around the M point due to the new nodal lines.}\label{surfstates}
\end{figure}

In order to analyze the nontrivial topological character of the nodal lines, we have computed the sum of the Wannier charge centers along a high symmetry path in the $k_x$-$k_y$ plane over the z-direction (WCC$_z(k_x,k_y)$) \cite{PhysRevB.84.075119,PhysRevB.89.115102}. This is equal (up to a factor of 2$\pi$) to the total Berry phase of the occupied Bloch functions accumulated along the $k_z$ direction, which has to be either 0 or $\pi$ for an arbitrary loop in the Brillouin zone. Thus, the WCC$_z(k_x,k_y)$ is expected to be quantized as either 0 or $1/2$ and a jump of $1/2$ will appear when passing through a projected nodal line.
At low pressure there is only one nodal line around the A point, which produces a jump of $1/2$ in WCC$_z(k_x,k_y)$ when crossing it (see Fig. \ref{wcc}a). This confirms the topological character of that nodal line, as already reported in Ref. \onlinecite{PhysRevB.93.201114}. In the high pressure regime new nodal lines emerge around the M points. Figure \ref{wcc}b shows that they produce a jump of $-1/2$ in WCC$_z(k_x,k_y)$. This is a proof of the nontrivial character also for these nodal lines. Note, however, that nodal lines around the M points have opposite sign to that around A. Since the M point has 3 branches it contributes three times. Adding contributions from all the nodal lines, at high pressures the system behaves as it would have two nodal lines, i.e. this phase transition has increased the amount of topological nodal lines by 1.

\begin{figure*}
  \centering
  \includegraphics[width=1.0\textwidth]
    {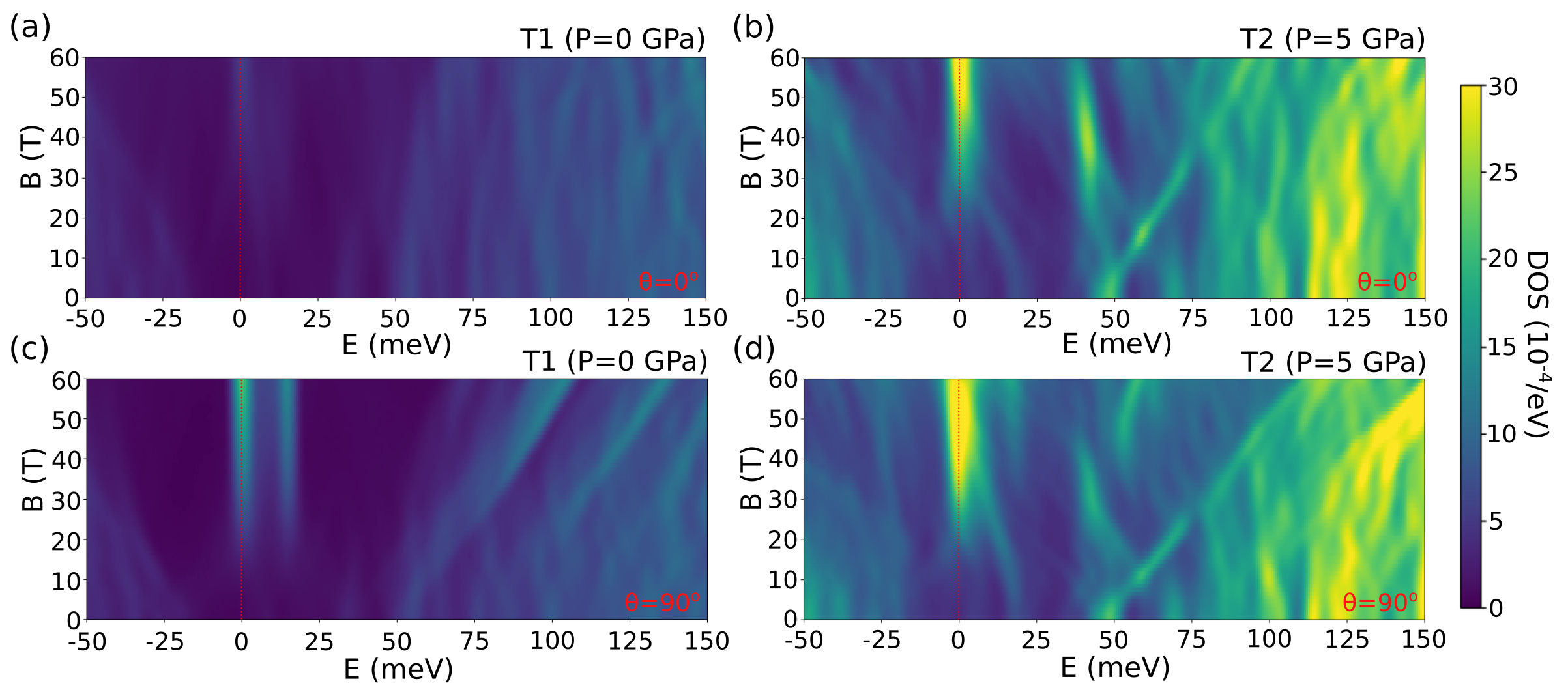}
     \caption{Colormap of the density of states as a function of an applied magnetic field B along the z-direction  (a,b) and perpendicular to it (c,d) for the T1 phase (a,c) and for the T2 phase (b,d). Magnetic oscillations and zero-energy modes are analyzed in the text. }\label{magosc}
\end{figure*}

It is important to remember that topological phase transitions are subject to energy considerations to take place as other phase transitions. In the present case, as we are neglecting SOC, the system is spin rotation invariant, so the symmetry protecting the nodal line at ambient pressure is the combined symmetry of inversion and time reversal symmetries. As this symmetry is not removed during the transition, as can be easily concluded from the fact that the initial nodal line remains unaltered, and the same combined symmetry protects the second nodal line appearing after $P=4$ GPa. Then, only energetic considerations are behind this topological phase transition.
This simple reasoning also allows to write a symmetry-allowed $k\cdot p$ model for the second nodal line around the $M$ point dictated by $PT$ symmetry, similar to the initial nodal line at ambient pressure\cite{PhysRevB.93.201114}:

\begin{eqnarray}\label{effHam}
H(\bm{k})&=&F(k_x-k_z)\sigma_2+\nonumber\\
&+&(A-B(k_x+k_z)^2-Ck^2_y)\sigma_3.
\end{eqnarray}

The effective $k\cdot p$ model in  Eq.(\ref{effHam}) captures the symmetry properties of the system at the $M_1$ point. Contrary to the case of the nodal line at the $A$ point, the $M$ points posses lower symmetries: $C_2$ symmetry ($k_y\rightarrow -k_y$) and an inversion symmetry ($k_x\rightarrow -k_x, k_z\rightarrow -k_z $). These symmetries do not fix the tilting angle of the plane containing the nodal lines around these points. In the present case, it is observed that this tilting angle is about $\pi/4$. The nodal lines around the other two $M$ points can be obtained from Eq.(\ref{effHam}) by $2\pi/3$ rotations.   Also, this model captures the peculiarity that the new nodal lines lie on a $\pi/4$ plane with respect the $\Gamma-M$ plane. We omit higher order terms in the model Eq.(\ref{effHam}) that do not qualitatively change the topological content of the new nodal lines, as they only introduce distortions  from the perfect elliptical shape of the nodal line.
The combined analysis of ab initio calculations and the tight binding model suggests that the parameter $A$  changes sign after the critical pressure, and this captures the features of the hopping parameter $t_{d_{xz}p_z}$ that controls the transition in the TB model. The parameters in the model (\ref{effHam}) can be fitted to the numerical TB calculation. We found that the Fermi velocity parameter $F$ is weakly dependent with the pressure as well. For $P=6$GPa, above the transition pressure, we find that $A\simeq66$meV, $F\simeq0.39 a$eVm, $C\simeq-1.60a^2$eVm$^2$, and $B\simeq 0.10a^2$eVm$^2$. All these parameters are written in terms of the lattice constant $a=4.652$\AA.

As mentioned above, one important feature of the topological NLSM is their drumhead-like surface states.
The TB Hamiltonian allows to compute them for each phase separately. Figure \ref{surfstates}a shows the energy states of a 50-layer structure stacked along the z-direction for the T1 phase, i.e. at low pressures. The colormap represents the expectation value of the position operator along the z-direction, in blue (red) the bottom (top) layer and in grey the bulk states.
We can see the crossings related to the topological nodal line that appears in the bulk. The edge states fill the space outside the projected nodal line around $\Gamma$. This behavior is a consequence of the 1D atomic chains with time-reversal and spatial inversion symmetries \cite{PhysRevB.93.201114}. 
In the high pressure T2 phase (Fig. \ref{surfstates}b) new crossings, related to the new topological nodal line, appear around the M point for the bulk states. In this phase, the surface states get modified by the presence of the new nodal line projected around M. As in the low pressure phase, inside the nodal line the surface states are not present.
Moreover, the dispersion of all surface states is increased by uniaxial pressure. In the T1 phase the bandwidth of the surface state is about 0.2 eV, while in the T2 phase it increases to more than 1 eV.
Note also that the surface states are highly sensitive to the kind of truncation imposed to the structure. Here we chose to terminate with a Sn-rich plane for our calculations in the z-direction, but different cuts can be used to obtain surface states inside or outside the projected nodal loop as explained in Ref. \cite{PhysRevB.93.201114}. Cuts in different directions are presented in the Supplemental material.

Magnetotransport measurements are the most extended experimental tests employed to analyze the topological character of NLSM through the analysis of the corresponding Landau level fan diagrams. It is also known that the same information can be obtained from quantum oscillations in the density of states (DOS) \cite{PhysRevLett.120.146602,PhysRevB.97.205107,PhysRevB.97.165118}.
Together with the Berry phase contribution to the phase of the quantum oscillations, non-trivial topological phases in NLSM are characterized by the appearance of zero-energy Landau levels \cite{PhysRevB.92.045126} in the presence of a magnetic field. The orbits associated to this zero-energy Landau level would be in turn responsible for the presence of a nontrivial Berry phase in the Landau level fan diagram.

Therefore, we have included a magnetic field (B) in our TB Hamiltonian through a Peierls flux with 80 unit cells in each direction. The kernel polynomial method \cite{RevModPhys.78.275} was used to compute the DOS as a function of the applied magnetic field (Fig. \ref{magosc}). All the calculations were well converged with respect to the input parameters.

In the low-pressure regime T1, there is only one nodal loop, lying in the  $OZ$ plane. When the magnetic field is applied perpendicular to the nodal line ($\theta=0$), no zero-energy modes are present since all the magnetic oscillations are trivial (Fig. \ref{magosc}a). However, when the field is applied perpendicular to the $OZ$ plane ($\theta=\pi/2$) the DOS associated to a zero-energy Landau level is observed (Fig. \ref{magosc}c). These results are in agreement and equivalent to those of a single nodal line lying in the $k_x$-$k_y$ plane produced by a cubic lattice \cite{PhysRevB.97.205107}. 

In the high-pressure phase T2, new nodal lines tilted $\pi/4$ with respect to the $OZ$-direction are present around the M points. Thus, when the magnetic field 
is applied along the $OZ$-direction ($\theta=0$) only the contribution from these new nodal lines is observed in the DOS at zero energy (Fig. \ref{magosc}b). The presence of this zero-energy mode is related with the nontrivial Berry phase that can be measured by the analysis of the quantum oscillations in magnetotransport.
When the magnetic field B is applied perpendicular to the $OZ$ plane ($\theta=\pi/2$), the original nodal line and the new ones contribute to the DOS at $E=0$. Since both zero-energy modes correspond to an extreme orbit that dominates the amplitude of the quantum oscillations, we expect a substantial increase in the magneto-oscillation amplitude (Fig. \ref{magosc}d).
We are focusing on the qualitative analysis of the pressure-induced topological phase transitions in NLSM, a fully quantitative analysis of the topological character of the quantum oscillations goes beyond the scope of the present work. However, we can obtain more quantitative results by focusing on the features of the zero-energy DOS of each phase.

\begin{figure}[!h]
  \centering
  \includegraphics[width=\columnwidth]%
    {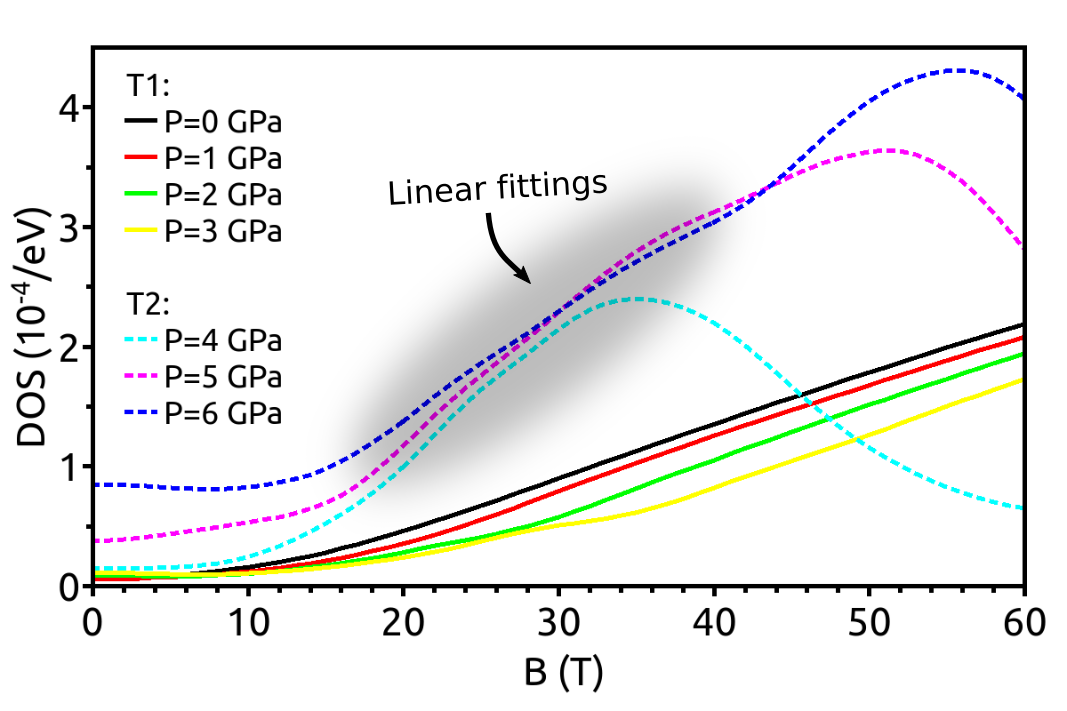}
     \caption{
     Zero-energy modes for different pressures as a function of the magnetic field. Linear fittings were performed for each phase. The ratio between slopes is 1.98, demonstrating the existence of a topological-to-topological phase transition (see text).}\label{zeromod}
\end{figure}

Figure \ref{zeromod} shows the DOS at zero energy as a function of B, for $\theta=\pi/2$ and for different pressure values. Both phases can be identified in this plot. While in the T1 phase there is only one nodal line contributing to the DOS at the Fermi level, in the T2 phase the emergence of 3 new nodal lines with opposite sign (as previously shown in the WCC$_z(k_x,k_y)$ analysis) makes the system behave as it would have 2 independent nodal lines. Therefore, the DOS at the Fermi level in the T2 phase is expected to be two times that in the T1 phase. To check whether this argument is correct, we have performed linear fittings of the DOS as a function of B.
While the dependence of the DOS with B in the T1 phase is almost linear since there is only 1 nodal line, the behavior in the T2 phase appears to be more complex as there are multiple nodal lines oriented along different directions. For this reason, we have restricted the fittings to a range of values in which the DOS shows a linear trend (shaded area in Fig. \ref{zeromod}). We have also averaged the slopes of each phase to reduce the uncertainty. 
We have obtained a ratio between the slope of phase T2 ($s_{T2}$) and  T1 ($s_{T1}$) of $s_{T2}/s_{T1}=1.98\simeq 2$. Therefore, this demonstrates a clear signature of the topological-to-topological phase transition.
Note that this smoking gun cannot be anticipated when SOC is included in the system. Calculations that go beyond the scope of the present work should be performed to answer that. However, it has been argued that for systems where SOC is sufficiently weak topological effects will still be observable in experiments \cite{PhysRevB.93.201114, ALLEN2018102, PhysRevLett.118.176402}.

\section{Conclusions} 

It has been widely reported, both theoretically and experimentally, that pressure can induce changes in the Fermi surface of different materials \cite{PhysRevB.85.174531, doi:10.1063/1.4816758, ZHANG2018280, Zhang2015}, and in some cases pressure has been found to modify the topological character of the compound. For instance, it has been shown that pressure can induce a transition to a nontrivial topological phase from a trivial one \cite{PhysRevB.89.035101, PhysRevB.84.245105, Zhou2020, PhysRevB.99.184109, PhysRevB.96.081112, Teshome2019, PhysRevB.96.241204}. The opposite case has been reported too. Pressure is able to induce a transition from topological to trivial in which topological objects such as nodal pairs \cite{PhysRevB.96.155205} or nodal lines are annihilated \cite{PhysRevB.99.085204}.
In our work, we report a different kind of transition (topological-to-topological) in BaSn$_2$ when SOC is neglected. Using BaSn$_2$ as a model, we have shown that pressure can increase the number of topological nodal lines in an already nontrivial compound.  

Our DFT calculations show that for uniaxial pressures along the z-direction greater than 4 GPa new nodal lines emerge in the Dirac NLSM BaSn$_2$. The topological character of these new nodal lines was studied with a model TB Hamiltonian that reproduces the DFT calculations as a function of pressure. The analysis of the WCC$_z(k_x,k_y)$, surface states and zero-energy modes when a magnetic field is applied, demonstrate that BaSn$_2$ in the absence of spin-orbit coupling undergoes a topological-to-topological phase transition at around 4GPa. Multiorbital character of the bands and crystal symmetry preservation have been found to be key factors to increase the number of topological nodal lines. 
In the present case, we propose a way to experimentally test this kind of topological-to-topological transition by noticing that the observation of the Berry phase in the fan diagram strongly depends on the tilting angle $\theta$ between the magnetic field and the nodal line \cite{PhysRevB.97.205107}. Thus, for a given tilting angle $\theta$ where the magneto-oscillations look trivial, the appearance of another nodal line with different tilting angle with respect to the magnetic field will generate topologically nontrivial magneto-oscillations. Therefore, providing a way to directly observe the topological transition described in the present work.

\section{DATA AVAILABILITY}
The data that support the findings of this study are available from the corresponding author upon reasonable request.

\section{ACKNOWLEDGEMENTS}

This work is supported by the MINECO of Spain through the project PGC2018-101334-B-C21. A.O.F. thanks MECD for the financial support received through the FPU grant FPU16/02572. A.C. acknowledges financial support through
MINECO/AEI/FEDER, UE Grant No. FIS2015-73454-
JIN and European Union structural funds, the Comunidad Aut\'onoma de Madrid (CAM) NMAT2D-CM Program (S2018-NMT-4511), and the Ram\'on y Cajal program through the grant RYC2018-023938-I. A. C. acknowledges useful discussions with Laszlo Oroszlany.

\section{AUTHOR CONTRIBUTIONS}
A.O.F., V.P., and A.C. conceived and designed the project. A.O.F. performed the DFT and TB model calculations, A.C. built the $k\cdot p$ model. All the authors contributed to the theoretical analysis and the manuscript writing.

\section{COMPETING INTERESTS}
The authors declare no competing interests.

\end{document}

\fi